# MDSA: Modified Distributed Storage Algorithm for Wireless Sensor Networks

Mohamed Labib Borham
Department of Computer Science
Faculty of Computers and Information
Helwan University

Mostafa-Sami Mostafa
Department of Computer Science
Faculty of Computers and Information
Helwan University

Hossam Eldeen Moustafa Shamardan
Department of Information TechnologyFaculty of Computers and Information
Helwan University

*Abstract*—In this paper, we propose a Modified distributed storage algorithm for wireless sensor networks (MDSA). Wireless Sensor Networks, as it is well known, suffer of power limitation, small memory capacity,and limited processing capabilities. Therefore, every node may disappear temporarily or permanently from the network due to many different reasons such as battery failure or physical damage. Since every node collects significant data about its region, it is important to find a methodology to recover these data in case of failure of the source node. Distributed storage algorithms provide reliable access to data through the redundancy spread over individual unreliable nodes. The proposed algorithm uses flooding to spread data over the network and unicasting to provide controlled data redundancy through the network. We evaluate the performance of the proposed algorithm through implementation and simulation. We show the results and the performance evaluation of the proposed algorithm.

## I. INTRODUCTION

Along with the industrial development and technological progress, monitoring the environment plays a very vital role. Many research works have been built-up on monitoring systems that can replace traditional systems in critical environments.

Wireless sensor networks (WSNs) are deployed to an area of interest to sense phenomena. Wireless sensor network is a type of ad-hoc networks that has the ability of sensing and processing data collected from the environment [1]. These networks are comprised of autonomous devices, called sensor nodes. Each sensor has a buffer which can be divided into small slots.Wireless Sensor Networks are recently applied in many environmental applications such as measuring temperature, humidity, salts or monitoring objects and others. WSNs applications have common task, which is environmental monitoring. This task is realized by using nodes to sense data from the environment and sends it to the base station. In all applications, we must take into account to use an efficient data collection approach.

Wireless sensor networks have many advantages but the main problem is the limitation of node's resources. Due to that fact, any node may fail to communicate with other nodes in the network or disappear from the network due to accidental events in harsh environments or due to battery depletion. In many applications, every sensor has important data thatareused to form the total overview about the environment so it is important to find an efficient method to recover the lost sensed data.

The Distributed Storage Algorithm (DSA-I) [2] was introduced as a model for recovering data from the failed nodes in WSNs. DSA-I algorithm depends on using flooding and multicasting to disseminate data packets over all network nodes. Flooding with multicasting cause a crowded messages through the network that rise power overhead, growth memory usage, increase-processing load on the nodes beside the increaseof both packet loss rate, and latency.

In this paper, we propose an unusual methodology which is called a Modified Distributed Storage Algorithm (MDSA) to overcome the previous algorithm DSA-I problems . MDSA algorithm depends on unicasting instead of multicasting to provide controlled-redundancy of data through the network. A consequence of these modifications, the number of created messages and the energy consumption are both reduced (See Table.1). Also, taking into account the buffer status whether it is empty or not. If it is full, it will cancel the operation to reduce processing load on the nodes. The proposed model is applied in large-scale wireless sensor network where these nodes are randomly distributed in the serviced environment.

The rest of this paper is organized as follows. In section 2, we present a short overview of distributed storage techniques and networking codes. In section 3, we introduce the algorithm design, consideration and assumptions of the proposed model. In section 4, we introduce our proposed MDSA algorithm. In section 5, we show a simulation of MDSA and comparison between the proposed algorithm and the initial algorithm DSA-I. In section 6, we explain our simulation results and performance evaluation. Finally, we conclude and present future work.

## II. RELATED WORK

Different types of networking codes allow every node in a network to achieve some computation. Therefore, each node is not used only to store data, but also has some processing capabilities to do some function or mixture when the data reach its destination; by that, the encoding process occurs inside the network and finally decoded at the final destination [3].

Using network codes provides reliable data access through redundancy over the nodes [4]. In [5], two schemes for maintaining redundancy-using erasure coding Maximum-Distance Separable (MDS) and Regenerating Codes (RC). In [6], it is used a decentralized implementation of Fountain codes that uses geographic routing and every node has to know its location. The motivation for using Fountain codes instead of using random linear codes is that Fountain codes need O( k ln k) decoding complexity but random linear codes and Reed-Solomon codes (RS) use O (k^3) decoding complexity where k is the number of data blocks to be encoded.

In [7], a technique is presented to increase data persistence in wireless sensor networks, which is called "Growth codes". This technique increases the amount of information that can be recovered at the sink node. "Growth codes" is a linear technique in which information is encoded in an online-distributed way with increasing degree. The authors defined persistence of a sensor network as "the fraction of data generated within the network that eventually reaches the sink node ".They showed that Growth codes can increase the amount of information that can be recovered at any storage node at any time period, whenever there is a failure in some other nodes. The motivation for their work is the positions of the nodes are not identified. In addition, a sensor node cannot know the position of other nodes. They assume around time of updating the nodes, meaning with increasing the time, the degree of a symbol is increasing. This is the idea behind growth degrees. They provide practical implementation of Growth codes and compare its performance with other codes.

In [7], the authors studied the question "how to retrieve historical data that the sensors have gathered even if some sensors are destroyed or disappeared from the network?" They analyzed techniques to increase "persistence" of sensed data in a randomly distributed wireless sensor network. They proposed two decentralized algorithms using Fountain codes to guarantee the persistence and reliability of cached data on unreliable sensors. They used random walks to disseminate data from a sensor (source) node to a set of other storage nodes. The first algorithm Exact Decentralized Fountain codes (EDFC) introduces lower overhead than naive random walk, while the second algorithm Approximate Decentralized Fountain Codes (ADFC) has a lower level of fault tolerance than the original centralized Fountain code, but consumes much lower dissemination cost.

In [7], the authors also proposed a novel decentralized implementation of Fountain codes in wireless sensor networks in an efficient and scalable fashion. The authors did not use routing tables to dissimilate data from one sensor to a set of sensors. The reason was the sensors did not have enough energy or memory to maintain routing table, which is scalable to the size of the network.

In [2 and 8], the authors presented a model for distributed storage algorithms for wireless sensor networks where K sensor nodes (sources) want to disseminate their data to N storage nodes with less computational complexity. They proposed two distributed storage algorithms that used flooding, Lt (Luby Transform) codes.Ltcodesare a special class of Fountain codes to guarantee the persistence of data and random walks to disseminate data between sensors. They also assumed in the first algorithm DSA-I that the total number or sources and storage nodes are known. The second algorithm DSA-II assumed that the total number or sources and storage nodes are not known. In other words, every node in the network can know only the number of neighboring nodes; also can estimate the number of sources and the total number of nodes.

### III. DESIGN CONSIDERATION AND ASSUMPTIONS

In this section, we show the model assumptions, we assume that all the nodes in the network are identical in all capabilities and act as sensing and storage node. All nodes are distributed randomly and uniformly in an environment, no node maintains routing or geographic tables. Every node can send a flooding message to the neighboring nodes; Also every node can discover the total number of neighbors by broadcasting a query message, and whoever replies to this message will be a neighbor of this node. All nodes have storage buffer. The buffer size is assumed 10 % of the network size. Every node prepares a packet with its ID, sensed data, hop counter, and a flag that is set to zero or one.

### IV. MDSA ALGORITHM

We will present a Modified Distributed Storage Algorithm (MDSA) for wireless sensor networks, where all nodes are able to sense and store data. The algorithm consists of two main operations:

A. Data separation and storage.

B. Data collection

*A. Data separation and storage*

Data separation and storage operations consist of three main phases: Network initialization, packet preparing and flooding phase, finally Data storage and Unicast. In the coming sections we will describe each phase in sequence.

*1) Network Initialization phase:*

Wireless sensor network applications are used in many environments that may be dynamic environment. In a dynamic environment, the location of the nodes is not fixed and for any environmental reason the nodes may be moved. In the proposed algorithm assumptions, we assume that at the beginning of running, the network topology is not known and the nodes do not maintain routing or geographic tables. Therefore, each node floods query message in its range, and every node that reacts is considered as a neighbor node. At the end of this phase, every node can count the total number of the neighboring nodes.

*2) Packet preparing and flooding phase:*

After the initialization phase, each node senses data about the required range, and then preparing one packet that is stored in the first slot in its buffer. This packet consists of four fields:

Packet= (ID, Sensed data, hop_count, Flag)

ID shows the sensed node ID, every node has a unique ID, and it is ordered according to its order of creation. The second field presents the sensed data from the environment and it is different in size according to the wireless sensor application purpose. The third field presents the hop count, which shows how many times the packet, will be traveled or transferred in the network; every node calculates its hop count according to the number of neighboring nodes. The hop_count is calculated according to the following equation:

$$hop\_count = \lfloor \text{The total number of network nodes} / \text{the total number of neighboring nodes} \rfloor.$$

The last field is a flag to show if the packet is new or old, zero for old data and one for updated packet. After preparing packets, each node floods its packet to all neighboring nodes. See Fig.1.

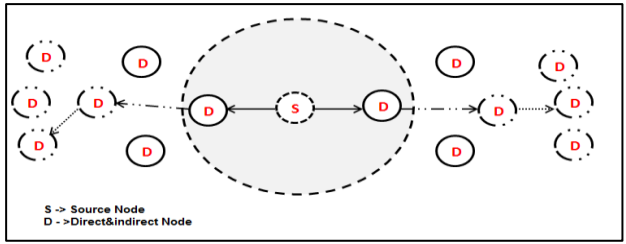

Figure.1 A WSN is randomly distributed in a field. A node **S** is source node and flood its data to its neighboring nodes D then D unicast its data to one of its neighbors and so on.

*3) Data storage and Unicast phase:*

After flooding of the packets from all the source nodes, the neighboring nodes store the received packet in its buffer if it has a location, and check the hop count if it is greater than zero, the received node will be unicasted the packet to one node from its neighboring nodes. This is repeated until the hop count equal zero the packet will be discarded. Fig.2. Show the MDSA Algorithm Flowchart where S is source node and I is intermediate node.

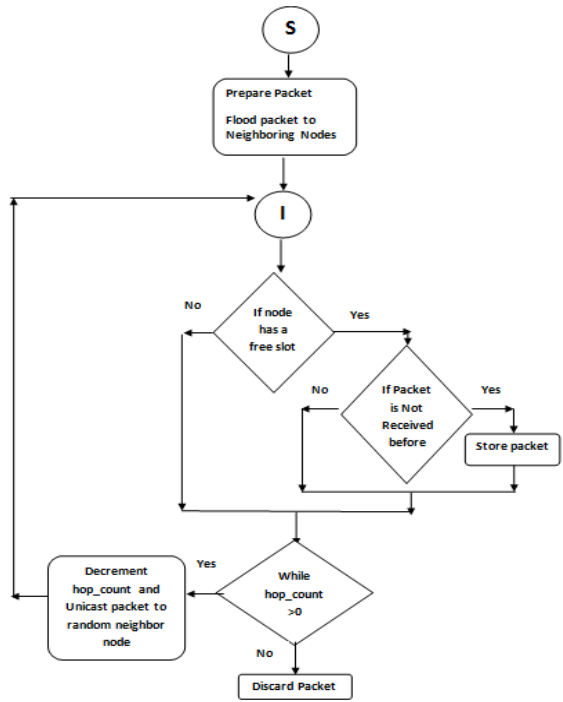

Figure.2 Modified Distributed Storage Algorithm Flowchart (MDSA)

*B. Data collection operations*

Data separation and storage operations provide data redundancy in the network so the stored data can be recovered by querying a number of living nodes from the network.

## V. MDSA SIMULATION AND COMPARISON

We simulate the proposed algorithm MDSA using well-known wireless sensor network simulator called OMNET++ 4. The simulation results show that the performance of the proposed algorithm is better than the previous algorithm DSA-I [2]. The comparisonmetricsare considering many factors, such as energy consumption, number of created messages, memory usage (see Fig.3), data dissemination and Recovering.

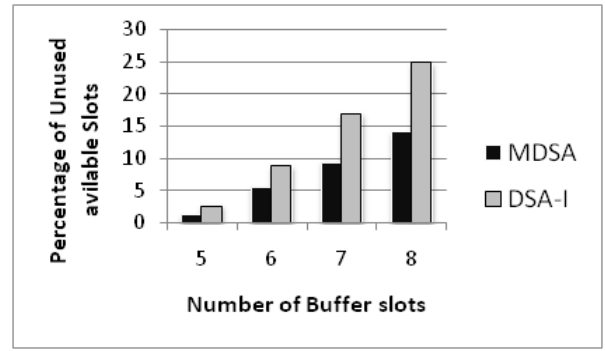

Figure.3 percentage of unused available buffer slots in MDSA and DSA-I algorithm

In Table.1, it is shown the practical comparison between the proposed algorithm MDSA and DSA-I [2] regarding to the number of created messages and percentage of unused buffer's slots with fixed number of nodes = 15, where M is the number of slots in every node's buffer.

TABLE I. COMPARISON BETWEEN MDSA AND DSA-I ALGORITHM

| Algorithm | Total Number of Nodes-N | No of Created Messages | Percentage of unused buffer's slots | | | |
|---|---|---|---|---|---|---|
| | | | M=5 | M=6 | M=7 | M=8 |
| DSA-I (old) | 15 | 29257 | 2.7% | 8.9% | 17.1% | 25% |
| MDSA (New) | 15 | 76 | 1.3% | 5.6% | 9.5% | 14.2% |

## VI. PERFORMANCE EVALUATION AND SIMULATION RESULTS

After simulating the Modified distributed storage algorithm using OMNET++ 4 simulator, the main performance metric to be investigated, is the average of the successful decoding percentage versus the decoding ratio.

We define the Average of successful decoding percentage as a percentage of recovered data to the total data in the network. In addition, we can define the decoding ratio as the number of nodes, which are queried, divided by the total number of network nodes.

Fig.4 to Fig.9 shows the effect of changing the total number of the network nodes. Moreover, we ran the experiments for 50,100,150,200,400,600 nodes with fixed ratio to the buffer size to be 10% of the network size.

We ran the experiment for many times reach to 30 times to evaluate the performance with various decoding ratios depending on the total number of nodes inside the network with incremental step = 0.1 , at every step we calculate the mean and the standard deviation to make sure that our results are closer to the actual implementation of the algorithm.

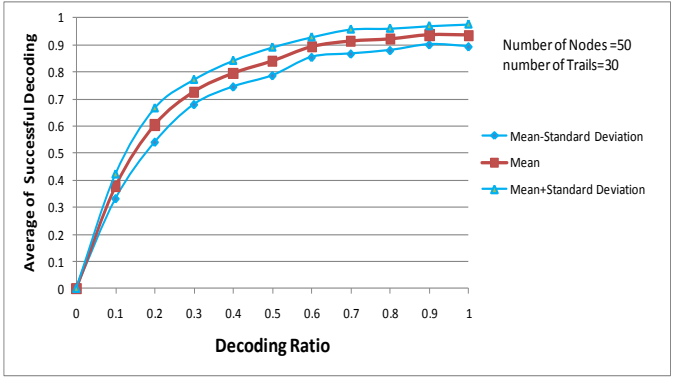

Figure.4.A WSN is randomly distributed in a field. The average of successful decoding ratio is shown for network size=50 nodes and buffer size= 5 slots with the MDSA algorithm

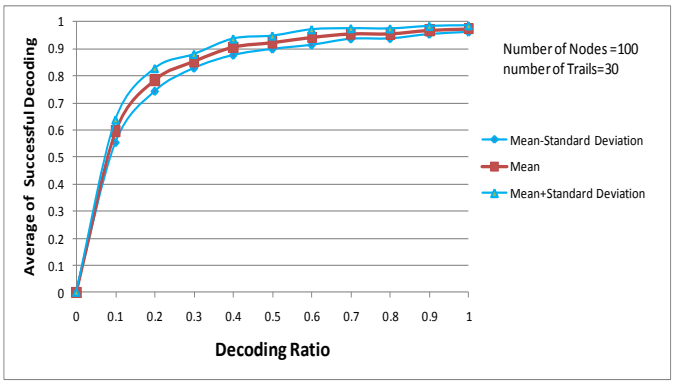

Figure.5.A WSN is randomly distributed in a field. The average of successful decoding ratio is shown for network size=100 nodes and buffer size= 10 slots with the MDSA algorithm

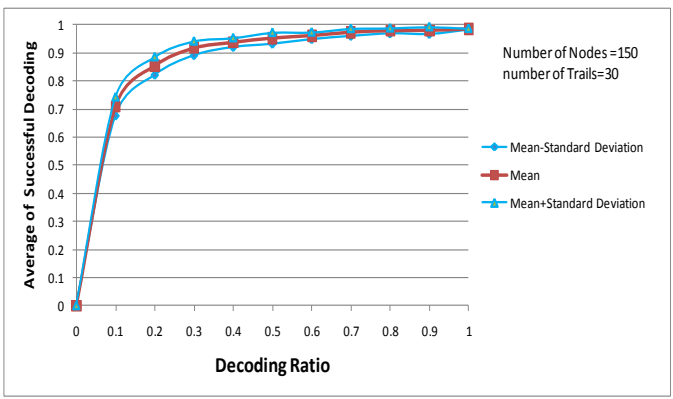

Figure.6.A WSN is randomly distributed in a field. The average of successful decoding ratio is shown for network size=150 nodes and buffer size= 15 slots with the MDSA algorithm

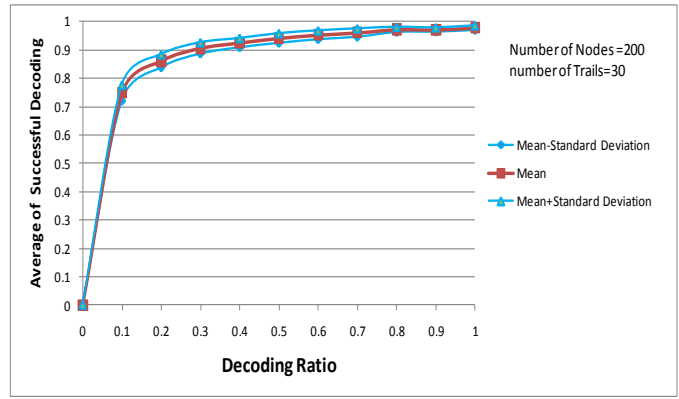

Figure.7.A WSN is randomly distributed in a field. The average of successful decoding ratio is shown for network size=200 nodes and buffer size= 20 slots with the MDSA algorithm

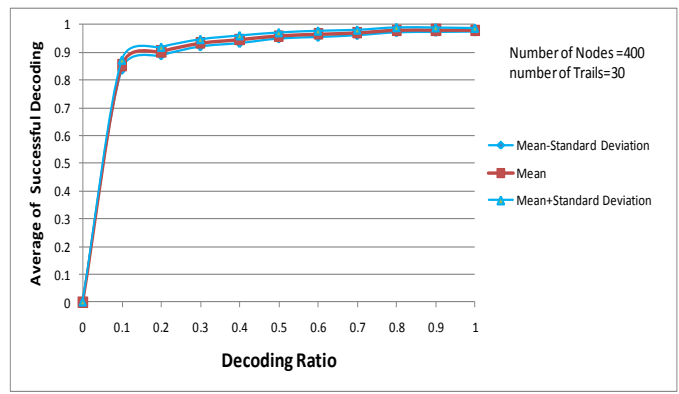

Figure.8.A WSN is randomly distributed in a field. The average of successful decoding ratio is shown for network size=400 nodes and buffer size= 40 slots with the MDSA algorithm

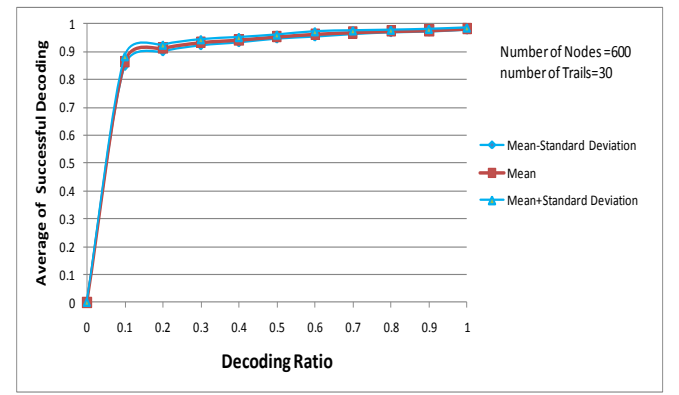

Figure.9.A WSN is randomly distributed in a field. The average of successful decoding ratio is shown for network size= 600 nodes and buffer size= 60 slots with the MDSA algorithm

VII. CONCLUSION AND FUTURE WORK

Modified distributed storage algorithm (MDSA) for wireless sensor networks is proposed in this paper to enhance data recovering for WSNs .This is achieved by disseminate sensed data throughout the networkusing flooding and unicasting to provide controlled-redundancy of data through the network.

The results and performance evaluation demonstrated that MDSAalgorithmis better on average of the successful decoding percentage according the same evaluationmetrics with the previous algorithm (DSA-I).

MDSA has shown better performance through increasing the network size.MDSA works to reduce unnecessary power consumption and memory usage when possible.Our future work will include practical and implementation aspects of this algorithm.

### ACKNOWLEDGMENT

I want to thank first my advisor,Prof.Dr.Mostafa-Sami M. Mostafa for his supervision, encouragement, and his guidance. Also, I would like to thank my supervisor Dr.HossamEldeenMoustafaShamardan, for his valuable advices and guiding me during this research work.

### REFERENCES

[1] I. F. Akyildiz, W. Su, Y. Sankarasubramaniam and E. Cayirci, "A survey on sensor networks," IEEE Communications Magazine, August 2002.

[2] S A Aly, H Darwish, M Youssef and M. Zidan,"Distributed Flooding-based storage algorithms for large-scale wireless sensor networks," In Proc. IEEE International Conference on Communication, Dresden, Germany, June 13-17, 2009.

[3] Philip A. Chou and Yunnan Wu," Network Coding for the Internet and Wireless Networks", Microsoft Research One Microsoft Way, Redmond, WA, 98052, June 2007.

[4] Zhenzhou Tang, Hongyu Wang, Qian Hu, and Long Hai, "How Network Coding Benefits Converge-Cast in Wireless Sensor Networks", In Proc. IEEE Vehicular Technology Conference (VTC 2012 Fall ),2012.

[5] A. G. Dimakis, P B Godfrey, M. Wainwright and K. Ramchandran, "The Benefits of Network coding for peer-to-peer storage," In Proc. Of 26th IEEE Infocom, Anchorage, AK, USA, May 6-12, 2007.

[6] Y. Lin, B. Liang, and B. Li. "Data persistence in large-scale sensor networks with decentralized fountain codes", In Proc. Of the 26th IEEE INFOCOM07, Anchorage, AK, May 6-12, 2007.

[7] A. Kamra, V. Misra, J. Feldman and D. Rubenstein, "Growth codes: Maximizing sensor network data persistence", In Proc. 2006 conference on Applications, Technologies, Architec- tures, and Protocols for Computer Communications, pp 255-266, Pisa, Italy, 2006.

[8] Salah A. Aly, Zhenning Kong and Emina Soljanin, "Fountain Codes Based Distributed Storage Algorithms for Large-scale Wireless Sensor Networks", International Conference on Information Processing in Sensor Networks. 2008.

AUTHORS PROFILE

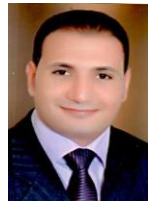

**Mohamed Labib Borham** is currently a Teaching Assistan, Faculty of Computer Science, Modern Sciences and Arts University (MSA, , 6 Octobe, , Egypt. He is a Masters Student at the Computer Science Department, Faculty of Computers and Information, Helwan University, Cairo, Egypt. He received her B.Sc. In Computer Science from Suez Canal University ,Ismalia. His current research interests include Wireless Sensor Networks Algorithms and applications,Artificial intelligence.

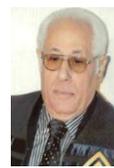

**Mostafa-Sami M. Mostafa** is currently a Professor of computer science, Faculty of Computers and Information, Helwan University, Cairo, Egypt. He worked as an Ex-Dean of faculty of Computers and Information Technology, MUST, Cairo. He worked also as an Ex-Dean of student affairs and Ex-Head of Computer Science Department, Faculty of Computers and Information, Helwan University, Cairo, Egypt. He is a Computer Engineer graduated 1967, MTC, Cairo, Egypt. He received his MSC 1977 and his PhD 1980 from University of Paul Sabatier, Toulouse, France. His research activities are in Software Engineering and Computer Networking. He is awarded supervising more than 80 Masters of Sc. And 18 PhDs in system modeling and design, software testing, middleware system development, real-time systems, computer graphics and animation, virtual reality, network security, wireless sensor networks and biomedical engineering.

**Hossam Eldeen Moustafa Shamardan** is currently a lecturer of Information Technology, Faculty of Computers and Information, Helwan University, Cairo, Egypt. He is an Electronics Engineer graduated 1991, Faculty of Engineering, Cairo University, Egypt. He received his MSC 1996 and his PhD 2005 from the University of Cairo, Giza , Egypt. His research activities are in Information security, Image processing, wireless sensor networks, and Computer Networking.